\documentstyle[12pt]{article}

\newcommand{\newc}{\newcommand}
\newc{\beq}{\begin{equation}}
\newc{\eeq}{\end{equation}}
\newc{\bea}{\begin{eqnarray}}
\newc{\eea}{\end{eqnarray}}
\newc{\kev}{\,\mbox{keV}}
\newc{\gev}{\,\mbox{GeV}}
\newc{\tev}{\,\mbox{TeV}}
\newc{\mev}{\,\mbox{MeV}}
\newc{\ev}{\,\mbox{eV}}
\newc{\gsim}{\lower.7ex\hbox{$\;\stackrel{\textstyle>}{\sim}\;$}}
\newc{\lsim}{\lower.7ex\hbox{$\;\stackrel{\textstyle<}{\sim}\;$}}
\newc{\mz}{m_Z}
\newc{\mpl}{M_{Pl}}
\newc{\obs}{{{\cal O}}}
\newc{\pb}{\,{\rm pb}}
\newc{\nn}{\nonumber}
\newc{\ie}{{\it i.e.}}
\newc{\eg}{{\it e.g.}}
\newc{\etal}{{\it et al.}}
\newc{\Z}{{\cal Z}}
\newc{\N}{{\cal N}}
\renewcommand{\bar}{\overline}
\def\NPB#1#2#3{Nucl.\ Phys.\ {\bf B#1} (19#2) #3}
\def\PLB#1#2#3{Phys.\ Lett.\ {\bf B#1} (19#2) #3}
\def\PLBold#1#2#3{Phys.\ Lett.\ {\bf#1B} (19#2) #3}
\def\PRD#1#2#3{Phys.\ Rev.\ {\bf D#1} (19#2) #3}
\def\PRL#1#2#3{Phys. Rev. Lett. {\bf#1} (19#2) #3}
\def\PRT#1#2#3{Phys.\ Rep.\ {\bf#1} (19#2) #3}

\def\MPL#1#2#3{Mod.\ Phys.\ Lett.\ {\bf #1} (19#2) #3}

\def\RMP#1#2#3{Rev.\ Mod.\ Phys.\  {\bf #1} (19#2) #3}

\catcode`@=11
\long\def\@caption#1[#2]#3{\par\addcontentsline{\csname
  ext@#1\endcsname}{#1}{\protect\numberline{\csname
  the#1\endcsname}{\ignorespaces #2}}\begingroup
    \small
    \@parboxrestore
    \@makecaption{\csname fnum@#1\endcsname}{\ignorespaces #3}\par
  \endgroup}
\catcode`@=12

\begin{document}
\begin{titlepage}
\begin{flushright}
{\rm
IASSNS-HEP-97-114\\
hep-ph/9710441\\
October 1997\\
}
\end{flushright}
\vskip 2cm
\begin{center}
{\Large\bf Implications of Generalized $Z-Z'$ Mixing}
\vskip 1cm
{\large
K.S.~Babu\footnote{Email: {\tt babu@ias.edu}},
Christopher Kolda\footnote{Email: {\tt kolda@ias.edu}} and 
John March-Russell\footnote{Email: {\tt jmr@ias.edu}}\\
}
\vskip 4pt
{\large\sl School of Natural Sciences,\\ Institute for Advanced Study,\\
Princeton, NJ, USA~~08540\\}
\end{center}
\vskip .5cm
\begin{abstract}
We discuss experimental
implications of extending the gauge structure of the Standard Model
to include an additional $U(1)$ interaction broken at or near the weak scale.
We work with the most general, renormalizable Lagrangian for the $SU(2)\times
U(1)\times U(1)$ sector, with emphasis on the phenomenon of gauge kinetic 
mixing between the two $U(1)$ gauge fields, and do not restrict ourselves
to any of the ``canonical'' $Z'$ models often discussed in the literature. 
Low-energy processes and $Z^0$-pole
precision measurements are specifically addressed.
\end{abstract}
\end{titlepage}
\setcounter{footnote}{0}
\setcounter{page}{1}
\setcounter{section}{0}
\setcounter{subsection}{0}
\setcounter{subsubsection}{0}


One of the simplest, and most well-motivated, extensions of the
Standard Model (SM) is the addition of an extra $U(1)$ gauge
factor to its $SU(3)\times SU(2)\times U(1)$ structure.
Traditionally~\cite{hr}, the most studied such extensions
were motivated by grand unified theories (GUT's) of rank higher
than that of the SM ($SO(10)$, $E_6$ or larger), or by
geometric compactifications of heterotic string models
which possessed a low-energy spectrum sharing many features with
$E_6$ GUT's.  Examples of the resulting $Z'$ gauge bosons include
the $\chi$ and $\psi$ models whose couplings to the SM fermions
are defined by the decompositions $SO(10)\to SU(5)\times U(1)_\chi$
and $E_6 \to SO(10)\times U(1)_\psi$.  However, it has recently become
clear that these models are not especially favored, and that a larger
class of $Z'$ models is more naturally considered.  

There are two
reasons for this change of view: first, even in ``traditional'' GUT-like
models, the phenomenon of kinetic mixing~\cite{holdom1,delaguila,bkm}
can significantly
shift the predicted couplings of the $Z'$ to SM states away from their
canonical values, as well as changing the relationship between
other SM observables~\cite{holdom2}; furthermore, kinetic mixing is in
general generated by renormalization group (RG) running down from the high
(\ie, GUT) scale to the weak scale~\cite{delaguila,bkm}.  Second, from the
string perspective, a much broader class of models with additional $U(1)$
factors now looks reasonable; this is due both to the construction of
many non-geometrical string models in weak coupling perturbation theory
(\eg, those arising from free-fermionic techniques) which share very little
resemblence to GUT-like models, and to the recent developements in 
strongly-coupled string theory which show that additional gauge factors can 
arise non-perturbatively.
It is also worth mentioning that string models naturally lead to non-zero
kinetic mixing as a threshold effect at the string scale~\cite{dkmr}.  In
either case the traditional parameterization in terms of the $U(1)'$ 
combinations in $E_6$ is inadequate.

Our intention in this paper is to explore the experimental consequences of the
most general $U(1)$ extensions of the SM.  Consistent
with the effective field theory philosophy, we study the full set
of additional renormalizable operators generated by the interactions of the 
$Z'$,
and allowed by SM and $U(1)'$ gauge symmetries. One expects that additional
{\sl matter} fields would also arise in extended gauge models in order to
cancel possible anomalies; however we will not study their effects here
as the spectrum and charges of extra matter would be highly model-dependent.

For a generic $Z'$, one usually find that direct experimental searches, via,
for example, Drell-Yan $Z'$ production at a $\bar pp$ collider, will provide
the strongest constraints on the existence of these new interactions. Such
searches have been considered many times both in the theoretical and 
experimental literatures. However there exist a number of low-energy processes
which are sensitive to extra gauge interactions, as well as $Z$-pole 
processes sensitive to $Z-Z'$ mixing. These processes can provide the most
useful mechanisms for searching for and/or measuring $Z'$ physics, 
particularly if the extra $Z'$ is either leptophobic~\cite{bkm,hadrophilic} 
or close in mass to 
the $Z$. It is on these indirect and mixing bounds that we will concentrate
in this work.

Finally, we wish to make note of previous related works which considered
many of these same issues in particular limits:
Kim, \etal~\cite{kim}\ on low-energy observables; Altarelli,
\etal~\cite{altarelli}\ on $Z^0$-pole observables; and Holdom~\cite{holdom2}
and Burgess, \etal~\cite{burgess}\ on electroweak radiative corrections.

\section{The Lagrangian}

The most general renormalizable Lagrangian
for the Standard Model with an additional 
$U(1)$ (denoted $U(1)'$) is given by (for convenience the QCD and 
scalar sectors have been omitted):
\bea
{\cal L}&=&{\cal L}_{{\rm SM}}+{\cal L}_{Z'}+{\cal L}_{{\rm mix}} \\
{\cal L}_{{\rm SM}}&=&-\frac{1}{4}\hat B_{\mu\nu}\hat B^{\mu\nu}
-\frac{1}{4}\hat W^a_{\mu\nu}\hat W^{a\mu\nu}+\frac{1}{2}\hat M_Z^2\hat Z_\mu 
\hat Z^\mu \\
& &{}-\hat e\sum_i\overline\psi_i\gamma^\mu\left(\frac{1}{\hat c_W}
(Y^i_LP_L+Y^i_RP_R)\hat B_\mu
+\frac{1}{\hat s_W}P_L\,T^a\cdot{\hat W^a_\mu}\right)\psi_i \nn \\
{\cal L}_{Z'}&=&-\frac{1}{4}\hat Z'_{\mu\nu}\hat Z'^{\mu\nu}
+ \frac{1}{2}\hat M_{Z'}^2 \hat Z'_\mu \hat Z'^\mu  -\frac{\hat g'}{2}
\sum_i\overline\psi_i\gamma^\mu (f^i_V-f^i_A\gamma^5)\psi_i \hat Z'_\mu\\ 
{\cal L}_{{\rm mix}}&=&{}-\frac{\sin\chi}{2}
\hat Z'_{\mu\nu}\hat B^{\mu\nu}+ \delta\hat M^2\, \hat Z'_\mu \hat Z^\mu
\eea
where: $\hat B_{\mu\nu},\hat W_{\mu\nu},\hat Z'_{\mu\nu}$ are the field 
strength tensors for $U(1)_Y$, $SU(2)_L$ and $U(1)'$ respectively;
$\psi_i$ are the fermion fields;
$f^i_V$ and $f^i_A$ are the vector and axial charges of the fermions under
$U(1)'$; $Y^i_L$ and $Y^i_R$ are the hypercharges of the left- and right-handed
components of the fermions with the normalization $Y^e_L=-\frac{1}{2}$
and $Y^e_R=-1$;
$\hat Z_\mu=\hat c_W W^3_\mu-\hat s_W \hat B_\mu$ 
is the usual $Z$-boson of the Standard Model (the usual photon, $\hat A_\mu$
is orthogonal to $\hat Z_\mu$); 
$\hat Z'$ is the boson of the new $U(1)'$; $\hat s_W$ is the sine of the 
weak angle; and $P_{L,R}=(1\mp\gamma^5)/2$. 
The mass terms
are assumed to come from spontaneous symmetry breaking via scalar
expectation values. 

The most noticable feature of the above Lagrangian is the presence of the
mixing terms in ${\cal L}_{{\rm mix}}$. The second (mass-mixing) term is 
familiar
and generally arises when the Higgs bosons of one group are also charged 
under the second group. The first term is less familiar and is allowed
for the case of two abelian groups only because $F_{\mu\nu}$ is 
gauge-invariant for abelian groups.
(As an aside, though, we note that the above
Lagrangian and all subsequent discussion do generalize to the case of new
electrically neutral gauge bosons coming from nonabelian gauge groups, 
except that $\chi=0$ there by gauge invariance.) Most previous analyses 
of extra $U(1)$'s have assumed $\chi=0$ at the weak scale even though it is
often generated by threshold corrections or via renormalization
group flow. It is the purpose of the present analysis to pay special attention
to the new contributions to physical processes that arise for $\chi\neq0$.

In order to discuss the physical implications of the new gauge boson,
it is necessary to work in the physical (or mass) eigenbasis for the
$Z-Z'$ system.
Going to the physical eigenbasis requires both diagonalizing the field
strength terms and the mass terms. This can be seen as a two-step process
in which we first diagonalize the field strengths via a $GL(2,R)$ 
transformation:
\beq
\left(\begin{array}{c} \hat B_\mu \\ \hat Z'_\mu \end{array}\right)=
\left(\begin{array}{cc} 1 & -\tan\chi \\ 0 & 1/\cos\chi \end{array}\right)
\left(\begin{array}{c} B_\mu \\ Z'_\mu \end{array}\right).
\eeq
Notationally, we express all parameters and fields in the original, mixed
basis as hatted, and those in the physical basis without hats.
So $(\hat B_\mu, \hat Z'_\mu)$ are the original $U(1)_Y$ and $U(1)'$ 
gauge fields with non-diagonal kinetic terms,
while $(B_\mu, Z'_\mu)$ have canonical gauge kinetic terms. The process
of diagonalization introduces new interactions among the gauge bosons and 
the matter
fields. Specifically, the $U(1)'$ charge of all fields is shifted by an amount
proportional to their hypercharge and $\sin\chi$; thus fields which were not 
charged
under $U(1)'$ now have some non-zero $U(1)'$ charge due to the non-orthogonal
rotation above.

Next
we go to the physical eigenbasis (via an $O(3)$ rotation) 
by diagonalizing the mass terms which arise
after both $U(1)'$-breaking and $SU(2)\times U(1)$-breaking. In the end,
one mass eigenstate is massless (the photon, $A_\mu$), 
while the other two (denoted $Z_{1,2}$) receive masses. In terms of
$(B_\mu, W^3_\mu, Z'_\mu)$, or alternatively $(\hat A_\mu, \hat Z_\mu,
\hat Z'_\mu)$, one finds:
\bea
A_\mu &=& \hat c_W B_\mu+\hat s_W W^3_\mu \nn \\
	&=& \hat A_\mu + \hat c_W \sin\chi \hat Z'_\mu \nn\\
Z_{1\mu}&=& \cos\xi (\hat c_W W^3_\mu-\hat s_W B_\mu)+\sin\xi Z'_\mu \\
	&=& \cos\xi(\hat Z_\mu-\hat s_W\sin\chi\hat Z'_\mu)+\sin\xi\cos\chi
	\hat Z'_\mu \nn \\
Z_{2\mu}&=& \cos\xi Z'_\mu - \sin\xi (\hat c_W W^3_\mu-\hat s_W B_\mu) \nn \\
	&=& \cos\xi\cos\chi\hat Z'_\mu-\sin\xi(\hat Z_\mu-\hat s_W\sin\chi
	\hat Z'_\mu)\nn
\eea
where~\footnote{This equation corrects Eq.~(27) of Ref.~\cite{bkm} in which
the last term in the denominator has the incorrect sign.}
\beq
\tan2\xi=\frac{-2\cos\chi(\delta\hat M^2+\hat M_Z^2 \hat s_W\sin\chi)}
{\hat M_{Z'}^2-\hat M_Z^2\cos^2\chi+\hat M_Z^2\hat s_W^2\sin^2\chi
+2\,\delta \hat M^2\,\hat s_W\sin\chi}.
\label {eq:xi}
\eeq

In order to make contact with
experiment, we must choose some definition for the couplings in terms of 
well-measured parameters. First we notice that the Lagrangian for the 
photon is the canonical one:
\beq
{\cal L}_{{\rm A}}=\hat e \overline\psi_i\gamma^\mu Q^i A_\mu \psi_i
\eeq
so we identify $\hat e=e$ where $e=\sqrt{4\pi\alpha}$ 
is the usual electric charge.
Given that $G_F$, $\alpha$ and $M_{Z_1}$ are the
best-measured parameters in the SM, we
then make the conventional choice to define the ``physical'' weak angle via:
\beq
s_W^2c_W^2=\frac{\pi\alpha(M_{Z_1})}{\sqrt{2} G_F M_{Z_1}^2}.
\label{eq:swcw}
\eeq
However, Eq.~(\ref{eq:swcw}) is also true with the replacements 
$s_W\to\hat s_W$, $c_W\to\hat c_W$ and $M_{Z_1}\to\hat M_Z$, leading to the  
identity $s_Wc_W M_{Z_1}=\hat s_W\hat c_W\hat M_Z$.
Keeping only to leading order~\footnote{It is often the case that 
$\xi^2(M_{Z_2}^2/
M_{Z_1}^2)\sim O(\xi)$ so we will keep it when working at $O(\xi)$.}
in $\xi$ and 
isolating the $Z_1$ interactions, the Lagrangian then takes the form:
\bea
{\cal L}_{Z_1}&=&-\frac{e}{2s_Wc_W}\left(1+\frac{\xi^2}{2}
\left(\frac{M_{Z_2}^2}{M_{Z_1}^2}-1\right)
+\xi s_W\tan\chi\right) \label{eq:lagold}\\
& &{}\times\overline\psi_i\gamma^\mu\left\lbrace
\left(T^i_3-2Q^i s_*^2
+\xi\tilde f^i_V\right) -\left(T^i_3+\xi 
\tilde f^i_A\right)\gamma^5\right\rbrace\psi_i Z_{1\mu} \nn
\eea
where we have used
\beq
\hat M_{Z}^2= M_{Z_1}^2\left\lbrace1+\sin^2\xi\left(\frac{M_{Z_2}^2}{M_{Z_1}^2}
-1\right)\right\rbrace
\eeq
and defined 
\bea
\tilde f^i_{V,A}&=&\frac{\hat g'c_Ws_W}{e\cos\chi} f^i_{V,A} \nn \\
s_*^2&\equiv&\sin^2\theta_* \label{eq:somedefs} \\
&=&s_W^2+c_W^2s_W\xi\tan\chi-\frac{c_W^2s_W^2}{c_W^2-
s_W^2}\left(\frac{M_{Z_2}^2}{M_{Z_1}^2}-1\right)\xi^2 \nn 
\eea
The last equation defines yet another weak angle which appears only in the 
vector interaction vertices.

This Lagrangian has a very familiar form and can be taken over 
directly to the effective Lagrangian formulation of the $S,T,U$ 
parameters~\cite{holdom2,burgess,peskin}. In that formulation, the 
model-independent (\ie, $g'\to0$) part of the corrected $Z_1$ interaction 
Lagrangian has the form:
\beq
{\cal L}_{Z_1}=-\frac{e}{2 s_W c_W}\left(1+\frac{\alpha T}{2}\right)\,
\overline\psi_i\gamma^\mu\left((T^i_3-2Q^i s_*^2)-T^i_3\gamma^5\right)
\psi_i Z_{1\mu}.
\label{eq:lagnew}
\eeq
where
\beq
s_*^2=s_W^2+\frac{1}{c_W^2-s_W^2}\left(\frac{1}{4}\alpha S-c_W^2 s_W^2
\alpha T\right)
\label{eq:sstar}
\eeq
Comparing Eqs.~(\ref{eq:lagold}) and (\ref{eq:somedefs}) to 
Eqs.~(\ref{eq:lagnew}) and
(\ref{eq:sstar}) we can identify the $Z-Z'$ contributions to $S,T$ to be:
\bea
\alpha S&=&4\xi c_W^2 s_W\tan\chi \\
\alpha T&=&\xi^2\left(\frac{M_{Z_2}^2}{M_{Z_1}^2}-1\right)+2\xi s_W
\tan\chi \nn
\eea
which hold to lowest order in $\xi$.
Several comments are in order: First, $Z-Z'$ mixing without kinetic mixing
(\ie, $\chi=0$) always shifts $T$ to larger values since the $Z_1$
mass will always be smaller than the pure SM $Z$ mass (assuming $\hat M_{Z'}
>\hat M_Z$). However in the presence
of kinetic mixing, $T$ can have either sign even though $M_{Z_1}$ is still 
smaller than $M_Z$. Second, non-zero $S$, of either sign, can be generated 
only in the presence of kinetic mixing. Third, although $U$ does not appear
in the above formulation, it can be explicitly calculated and one finds it
remains zero to leading order in $\xi$.

The $Z_1\overline\psi\psi$ interaction Lagrangian can then be given in final 
form:
\beq
{\cal L}_{Z_1}=-\frac{e}{2s_Wc_W}\left(1+\frac{\alpha T}{2}\right) 
\overline\psi_i\gamma^\mu\left\lbrace \left(g^i_V+\xi\tilde f^i_V\right)
-\left(g^i_A+\xi\tilde f^i_A\right)\gamma^5\right\rbrace\psi_i Z_{1\mu}
\eeq
with the following identifications: 
\bea
g^i_A&=&T^i_3 \nn \\
g^i_V&=&T^i_3-2Q^i s_*^2 
\eea
and $s_*^2$ as defined in Eq.~(\ref{eq:somedefs}). 
The same procedure can also
produce the $Z_2\overline\psi\psi$ interaction Lagrangian:
\beq
{\cal L}_{Z_2}=-\frac{e}{2s_Wc_W}\overline\psi_i\gamma^\mu\left\lbrace
\left(h^i_V-g^i_V\xi\right)-\left(h^i_A-g^i_A\xi\right)
\gamma^5\right\rbrace\psi_i Z_{2\mu} 
\eeq
with the following additional definitions:
\bea
h^i_V&=&\tilde f^i_V+\tilde s(T^i_3-2Q^i)\tan\chi \nn \\
h^i_A&=&\tilde f^i_A+\tilde s T^i_3\tan\chi \\
\tilde s&\equiv&\sin\tilde{\theta} \nn \\
&=&s_W+\frac{s_W^3}{c_W^2-s_W^2}\left(
\frac{1}{4 c_W^2}\alpha S-\frac{1}{2}\alpha T\right) \nn
\eea
where the last equation defines yet another weak angle.
If $\chi$ is very small, then the shift $\tilde s^2-s_W^2$ is higher order in 
small
$\chi$ and $\xi$; however, there is no reason given current data to exclude 
the possibility of small $\xi$ but large $\chi$, since $\chi$ always 
appears suppressed by $\xi$ in the $Z_1$ interaction Lagrangian.

\section{Observables}

There are two sets of constraints on the existence of a $Z'$ which
will be considered here: precision measurements of neutral current processes 
at low energies, and $Z$-pole constraints on $Z-Z'$ mixing. In 
principle, one usually expects other new states to appear at the same 
scale as the $Z'$, including its symmetry-breaking sector and any additional
fermions necessary for anomaly cancellation. However, because these states are
highly model-dependent, we will not include their effects in the expressions
that follow.

\subsection{Low-energy observables} 

At energies far below $M_Z$, the effects of an additional gauge field with mass
$M_{Z'}\sim M_Z$ can easily be comparable to those of the electroweak $Z$.
Therefore in processes sensitive to off-shell $Z$-exchange, signals for 
$Z'$-exchange may also be constrained. We consider several such processes in
this section. Earlier discussions of these processes can be found in 
Refs.~\cite{kim,apv}.

Whether or not the new gauge interactions are parity violating, 
stringent constraints can 
arise from atomic parity violation (APV) experiments. 
At low energies, the effective Lagrangian for electron-quark interactions
can be written:
\beq
{\cal L}_{{\rm eff}}=\frac{G_F}{\sqrt{2}} \sum_{q=u,d} \left\lbrace 
C_{1q} (\bar e \gamma_\mu
\gamma^5 e)(\bar q \gamma^\mu q) + C_{2q} (\bar e \gamma_\mu e)
(\bar q \gamma^\mu \gamma^5 q) \right\rbrace
\eeq
where $C_1$ and $C_2$ are given in terms of the underlying physics by
\bea
C_{1q}&=&2(1+\alpha T)(g^e_A+\xi\tilde f^e_A)(g^q_V+\xi\tilde f^q_V)
+2r(h^e_A-\xi g^e_A)(h^q_V-\xi g^q_V) \\
C_{2q}&=&2(1+\alpha T)(g^e_V+\xi\tilde f^e_V)(g^q_A+\xi\tilde f^q_A)
+2r(h^e_V-\xi g^e_V)(h^q_A-\xi g^q_A) \nn
\eea
with $r=(M_{Z_1}/M_{Z_2})^2$. The second ($r$-dependent) term come from
$Z_2$ exchange and is often non-negligible given that $\xi\sim(M_{Z_1}/
M_{Z_2})^2$ in many cases. 

One then defines the ``weak charge,'' $Q_W$, of an atom:
\beq
Q_W=-2[C_{1u}(2\Z+\N)+C_{1d}(\Z+2\N)]
\eeq
where $\Z$ ($\N$) is the number of protons (neutrons) in the atom.
We can express the effects of $Z-Z'$ mixing as a shift in $Q_W$ away from the
(one-loop corrected) SM prediction:
\bea
\Delta Q_W&\simeq&-\frac{\Z}{c_W^2-s_W^2}\alpha S
+\left(Q_W+\frac{4c_W^2s_W^2\Z}
{c_W^2-s_W^2}\right)\alpha T \nn \\
& &{}-4(2\Z+\N)\left\lbrace
\xi g^e_A(\tilde f^u_V-rh^u_V)+\xi g^u_V(\tilde f^e_A-rh^e_A)+rh^e_Ah^u_V
\right\rbrace \\
& &{}-4(\Z+2\N)
\left\lbrace u\leftrightarrow d \right\rbrace \nn
\eea
to lowest order in $\xi$.

The corresponding $C_{2q}$ couplings are measured in combination with the
$C_{1q}$ couplings in polarized electron-nucleon scattering experiments.
The left-right asymmetry, $A_{LR}$, can be expressed in the quark model
as:
\beq
\frac{A}{{\cal Q}^2}=a_1+a_2\frac{1-(1-y)^2}{1+(1-y)^2} 
\eeq
where ${\cal Q}^2>0$ is the momentum transfer in the scattering, and 
$y$ is the fractional energy loss of the electrons in the nucleon rest
frame. The $a_i$ are defined via:
\bea
a_1=\frac{3G_F}{5\sqrt{2}\pi\alpha}\left(C_{1u}-\frac{1}{2}C_{1d}\right) \nn \\
a_2=\frac{3G_F}{5\sqrt{2}\pi\alpha}\left(C_{2u}-\frac{1}{2}C_{2d}\right).
\eea
We can again expand $\Delta a_1$ and $\Delta a_2$ to lowest order in $\xi$,
keeping all orders in $r$ and $\chi$:
\bea
\Delta a_1&=&a_1\alpha T+\frac{3G_F}{5\sqrt{2}\pi\alpha}\left\lbrace
2\xi\left(\frac{3}{4}-\frac{5}{3}s_W^2\right)(\tilde f^e_A+rh^e_A)
-\xi\left(\tilde f^u_V-\frac{1}{2}
\tilde f^d_V\right)\right. \nn \\
& &\left.{}+2r\left(h^e_A-\frac{1}{2}\xi\right)\left(h^u_V-\frac{1}{2}h^d_V
\right)
+\frac{5}{3(c_W^2-s_W^2)}\left(\frac{1}{4}\alpha S-c_W^2s_W^2\alpha T\right)
\right\rbrace \\
\Delta a_2&=&a_2\alpha T+\frac{3G_F}{5\sqrt{2}\pi\alpha}\left\lbrace
\frac{3}{2}\xi(\tilde f^e_V+rh^e_V)+2\xi\left(2s_W^2-\frac{1}{2}\right)
\left(\tilde f^u_A-\frac{1}{2}\tilde f^d_A\right)\right. \nn \\
& &\left.{}+2r\left(h^e_V+\left(2s_W^2-\frac{1}{2}\right)\xi\right)
\left(h^u_A-\frac{1}{2}h^d_A\right)\right\rbrace.
\eea

Stringent limits on $Z-Z'$ mixing also arise from neutrino-hadron 
scattering experiments. These experiments parametrize their results in
terms of the effective Lagrangian:
\beq
{\cal L}_{\nu{\rm Hadron}}=-\frac{G_F}{\sqrt{2}}\bar\nu\gamma^\mu
(1-\gamma^5)\nu \sum_q\left[\epsilon_L(q)\bar q\gamma_\mu(1-\gamma^5)
q+\epsilon_R(q)\bar q\gamma_\mu(1+\gamma^5)q\right].
\eeq
The $\epsilon$-parameters can be expressed as:
\bea
\epsilon_{L,R}(q)&=&\frac{1}{2}(1+\alpha T)\left\lbrace(g^q_V\pm g^q_A)
[1+\xi(\tilde f^\nu_V\pm\tilde f^\nu_A)]+\xi(\tilde f^q_V\pm\tilde f^q_A)
\right\rbrace \\
& &{}+\frac{r}{2}\left\lbrace (h^q_V\pm h^q_A)(h^\nu_V\pm h^\nu_A)
-\xi(g^q_V\pm g^q_A)(h^\nu_V\pm h^\nu_A) - \xi(h^q_V\pm h^q_A)
\right\rbrace. \nn
\eea
This can be expanded, keeping only terms linear in $\xi$, to determine a
shift $\Delta\epsilon_{L,R}(q)$, but it is trivial to do and we will abstain
from showing it here.

\subsection{$Z$-pole observables} 

Electroweak measurements made at LEP and SLC while
sitting on the $Z$-resonance have greatly suppressed sensitivities to 
$Z'$ physics {\it except} through the mixing with the $Z$. 
This is because the production rate for $Z_1\equiv Z^0$ would overwhelm
most $Z_2$ direct production signals, and at the pole, $Z-Z'$ interference
effects vanish. (We ignore in this section the
possibility that the $Z_2$ could be nearly degenerate with the usual $Z_1$.)

Constraints on the allowed mixing angle and $U(1)'$ charges arise by
fitting all data simultaneously to the {\it ansatz} of $Z-Z'$ mixing.
For any observable, $\obs$, 
the shift in that observable, $\Delta\obs$, can be expressed (following
the procedure of Refs.~\cite{altarelli,burgess}) as:
\beq
\frac{\Delta\obs}{\obs}={\cal A}^{S}_\obs\,\alpha S+{\cal A}^{T}_\obs\,\alpha T
+\xi\sum_i {\cal B}^{(i)}_\obs \tilde f^i
\label{eq:lep}
\eeq
where $i$ runs over the independent $Z'\bar\psi\psi$ 
couplings. We have of course assumed that the shifts on the right-hand
side of Eq.~(\ref{eq:lep}) are small so that the equation can be linearized.

If the $U(1)'$ charges are generation-dependent, there
exist severe constraints in the first two generations coming
from precision measurements such as the
$K_L-K_S$ mass splitting and $B(\mu\to 3e)$ owing to the lack
of GIM suppression in the $Z'$ interactions; however, constraints
on a $Z'$ which couples only to the third generation
are somewhat weaker. In any case, per generation there are
only five independent $Z'\bar\psi\psi$ couplings; we can choose them to be
$\tilde f^u_V, \tilde f^u_A, \tilde f^d_V, \tilde f^e_V, \tilde f^e_A$. 
All other couplings can be
determined in terms of these, {\it e.g.}, $\tilde f^\nu_V=(\tilde f^e_V
+\tilde f^e_A)/2$, 
$\tilde f^u_A=(\tilde f^e_V+\tilde f^e_A)/2$ and $\tilde f^d_A=\tilde f^u_V
+\tilde f^u_A-\tilde f^d_V$. Thus 
in Eq.~(\ref{eq:lep}), the variable $i$ runs over these 5 (per generation) 
independent variables. 

The ${\cal A}$ and ${\cal B}$ coefficients are given numerically
in Table~\ref{table}\ where we assume generation-independent $Z'$ interactions.
To lowest order, the coefficients in the table 
depend only on the measured SM parameters.
The observables we consider here include those measured directly at
LEP/SLC: $\Gamma_Z$, $R_\ell=\Gamma_{\rm had}/\Gamma_{\ell^+\ell^-}$,
$\sigma_{h}=12\pi\Gamma_e\Gamma_{{\rm had}}/M_{Z_1}^2\Gamma_Z^2$,
$R_b=\Gamma_{b}/\Gamma_{{\rm had}}$ and $R_c=\Gamma_c/\Gamma_{{\rm had}}$.
We also include the parameters $\bar A_e$, $\bar A_b$ and $\bar A_c$ which
are extracted from the corrected asymmetries: $A^{(0,f)}_{FB}=\frac{3}{4}
\bar A_e\bar A_f$ and $A^0_{LR}=\bar A_e$. 

\begin{table}
\centering
\begin{tabular}{cccccccc}
\hline \hline
$\obs$ & ${\cal A}^S_\obs$ & ${\cal A}^T_\obs$ 
& ${\cal B}^{Vu}_\obs$ & ${\cal B}^{Au}_\obs$ & ${\cal B}^{Vd}_\obs$ 
& ${\cal B}^{Ve}_\obs$ & ${\cal B}^{Ae}_\obs$ \\ \hline
$\Gamma_Z$ & -0.49 & 1.35 & -0.89 & -0.40 & 0.37 & 0.37 & 0 \\
$R_\ell$ & -0.39 & 0.28 & -1.3 & -0.56 & 0.52 & 0.30 & 4.0 \\
$\sigma_h$ & 0.046 & -0.033 & 0.50 & 0.22 & -0.21 & -1.0 & -4.0 \\
$R_b$ & 0.085 & -0.061 & -1.4 & -2.1 & 0.29 & 0 & 0 \\
$R_c$ & -0.16 & 0.12 & 2.7 & 4.1 & -0.59 & 0 & 0 \\
$\bar A_e$ & -24.9 & 17.7 & 0 & 0 & 0 & -26.7 & 2.0 \\
$\bar A_b$ & -0.32 & 0.23 & 0.71 & 0.71 & -1.73 & 0 & 0 \\
$\bar A_c$ & -2.42 & 1.72 & 3.89 & -1.49 & 0 & 0 & 0 \\
$M_W^2$ & -0.93 & 1.43 & 0 & 0 & 0 & 0 & 0 \\ \hline\hline
\end{tabular}
\caption{Strength of dependence of observables on parameters of $Z_1$ 
Lagrangian. See Eq.~(\protect\ref{eq:lep}).}
\label{table}
\end{table}

\section{Conclusions}

In this paper we have presented general expressions for the shifts in precision
measurements at low energies and at the $Z^0$-pole due to the presence of
an additional $U(1)$ interaction. 
In doing so, we have always used the most general 
parametrization of an $SU(2)\times U(1)\times U(1)'$ model, including gauge
kinetic mixing. We have made no expansions in small parameters apart from
the assumption that the mixing angle $\xi$ is small; 
in particular the expressions herein 
do not assume a large mass hierarchy between the $Z$ and $Z'$.

We have not
attempted to fit the present data to any particular model because the
Standard Model as is works quite nicely.
However, if deviations
from the Standard Model begin to appear and/or direct observations of a
new gauge boson are made, such fits will play an important part in 
extracting the physics of the new interactions. In particular, it would
be important 
to extract from the experimental data a measurement of the couplings 
$f^i_{V,A}$ and $\chi$ independently. This is a considerable challenge without 
foreknowledge of the underlying gauge structure; the question
of how to go about doing this is currently under study~\cite{bdkm}.

\section*{Acknowledgements}
This research was supported in part by US Department of Energy contract 
\#DE-FG02-90ER40542,
the Alfred P.~Sloan Foundation and the W.M.~Keck Foundation. CK also wishes
to acknowledge the generous support of Helen and Martin Chooljian.
Our gratitude goes out to the Theory Group
at the ICTP, Trieste for their hospitality during the
extended workshop on highlights in astroparticle physics, the
Aspen Center for Physics, and
the Department of Theoretical Physics, Oxford University where
parts of this work were completed. Finally we would like to thank 
C.~Carone and T.~Trippe for
encouraging us to put these results together in a concise form.

\end{document}